\documentstyle[12pt,oldlfont]{article}
\newcommand{\n}{\hspace*{-2.5mm}}
\newcommand{\simgt}{\,\rlap{\lower 3.5 pt \hbox{$\mathchar \sim$}} \raise 1pt
 \hbox {$>$}\,}
\newcommand{\simlt}{\,\rlap{\lower 3.5 pt \hbox{$\mathchar \sim$}} \raise 1pt
 \hbox {$<$}\,}

\begin{document}
\title{\vskip-3cm{\baselineskip14pt
\centerline{\normalsize DESY 96--001\hfill ISSN 0418--9833}
\centerline{\normalsize MPI/PhT/96--002\hfill}
\centerline{\normalsize hep--ph/9601278\hfill}
\centerline{\normalsize January 1996\hfill}}
\vskip1.5cm
Inclusive Hadron Production in Photon-Photon Collisions at Next-to-Leading
Order}
\author{J. Binnewies$^1$, B.A. Kniehl$^2$, G. Kramer$^1$\\
$^1$ II. Institut f\"ur Theoretische Physik\thanks{Supported
by Bundesministerium f\"ur Forschung und Technologie, Bonn, Germany,
under Contract 05~6~HH~93P~(5),
and by EEC Program {\it Human Capital and Mobility} through Network
{\it Physics at High Energy Colliders} under Contract
CHRX--CT93--0357 (DG12 COMA).},
Universit\"at Hamburg\\
Luruper Chaussee 149, 22761 Hamburg, Germany\\
$^2$ Max-Planck-Institut f\"ur Physik, Werner-Heisenberg-Institut\\
F\"ohringer Ring 6, 80805 Munich, Germany}
\date{}
\maketitle
\begin{abstract}
We study inclusive charged-hadron production in collisions of quasireal 
photons at next-to-leading order (NLO) in the QCD-improved parton model, using
fragmentation functions recently extracted from PEP and LEP1 data of $e^+e^-$
annihilation.
We consistently superimpose the direct (DD), single-resolved (DR), and
double-resolved (RR) $\gamma\gamma$ channels.
We consider photon spectra generated by electromagnetic bremsstrahlung and/or
beamstrahlung off colliding $e^+$ and $e^-$ beams as well as those which
result from backscattering of laser light off such beams.
First, we revisit existing single-tag data taken by TASSO at PETRA and by
MARK~II at PEP (with $e^+e^-$ energy $\sqrt S\approx30$~GeV) and confront them
with our NLO calculations imposing the respective experimental cuts.
We also make comparisons with the neutral-kaon to charged-hadron ratio
measured by MARK~II.
Then, we present NLO predictions for LEP2, a next-generation $e^+e^-$ linear
collider (NLC) in the TESLA design with $\sqrt S=500$~GeV, and a Compton
collider obtained by converting a 500-GeV NLC.
We analyze transverse-momentum and rapidity spectra with regard to
the scale dependence,
the interplay of the DD, DR, and RR components,
the sensitivity to the gluon density inside the resolved photon,
and the influence of gluon fragmentation.
It turns out that the inclusive measurement of small-$p_T$ hadrons at a
Compton collider would greatly constrain the gluon density of the photon and
the gluon fragmentation function.

\end{abstract}
\newpage

\section{Introduction}

The inclusive production of single hadrons in fixed-target and colliding-beam
experiments has been one of the most important testing grounds for the
QCD-improved parton model.
In contrast to the collective observation of jets of hadrons, the detection
of single mesons and baryons also allows one to study fragmentation,
{\it i.e.}, the mechanism of how partons (quarks, gluons, and photons) turn
into hadronic matter.
In the framework of the QCD-improved parton model, the cross section of 
inclusive single-hadron production is described as a convolution of the
parton-parton scattering cross sections with the parton density functions 
(PDF's) of the initial-state particles and the fragmentation functions (FF's),
which characterize the transition of the partons that come out of the hard
scattering to the hadrons that finally hit the detector.
The factorization theorem \cite{facth} ensures that the PDF's and FF's are
universal and that only the partonic cross sections change when different
processes are considered.
While the partonic cross sections may be perturbatively calculated from the
QCD Lagrangian, this is not yet possible for the PDF's and the FF's of hadrons
with masses smaller than or comparable to the asymptotic scale parameter,
$\Lambda$, and one has to determine them by fitting experimental data.
The PDF's of protons and photons, which are needed to describe, {\it e.g.},
$p{\bar p}$, $ep$, and $\gamma\gamma$ reactions,
are already highly constrained by measurements of deep-inelastic $ep$ and
$e\gamma$ processes.

The most direct way to obtain information on the FF's of hadrons is to analyze
their energy spectrum measured in $e^+e^-$ annihilation, where the theoretical
predictions are not obscured by additional nonperturbative input, {\it e.g.},
in the form of PDF's for the incoming particles.
After the extraction of leading-order (LO) FF's for pions and kaons from
low-energy data of $e^+e^-$ annihilation and muon-nucleon deep-inelastic
scattering in the late 70's and early 80's \cite{bai}, there had long been no
progress in this field.
Some time ago, new next-to-leading-order (NLO) FF sets for charged and neutral
pions and kaons as well as for eta mesons were fitted to $e^+e^-$ data
generated with well-established Monte Carlo (MC) programs, which are
fine-tuned so as to describe well a broad selection of experimental data
\cite{chi}.
An alternative approach is to directly fit to experimental data.
We took this avenue for charged pions and kaons \cite{bkk1,bkk2} and neutral
kaons \cite{bkk3} by analyzing SLAC-PEP and recent CERN-LEP1 data.
The assumption that the $s$, $c$, and $b$ ($d$, $c$, and $b$) quarks fragment
into charged pions (kaons) in the same way, which we had to make in
Ref.~\cite{bkk1}, could be discarded in Ref.~\cite{bkk2}, thanks to the advent
of accurate data from the ALEPH Collaboration at LEP1 \cite{alephpre},
in which the fragmentation of gluons, $b$ quarks, and light flavours into
charged hadrons was distinguished.
In the meantime, these data have become available in published form
\cite{alephpub}, where the fragmentation of quarks into charged hadrons is
also reported for an enriched $c$-quark sample.
This latter information was not at our disposal at the time of our analysis
\cite{bkk2}.
However, we verified that our $c$-quark FF's for charged hadrons are
approximately in agreement with this new measurement.
In order to probe the scaling violation predicted by the Altarelli-Parisi
equations \cite{ap}, we made comparisons with other $e^+e^-$ data collected at
different centre-of-mass (CM) energies.
To test the validity of the factorization theorem for fragmentation, we also
confronted measurements of inclusive single-hadron production by H1 \cite{h1}
and ZEUS \cite{zeus} at the DESY $ep$ collider HERA and by UA1 \cite{ua1} at
the CERN $Sp\bar pS$ collider with the corresponding theoretical predictions
based on our FF's \cite{bkk2,bkk3,bor}.\footnote{The NLO results shown in
Ref.~\cite{ua1} were taken from Ref.~\cite{bor}, where the FF's of
Ref.~\cite{bkk1} were employed.}
In all cases, we found good agreement.
Existing data on inclusive single-hadron production in collisions of quasireal
photons offer yet another opportunity to quantitatively check the
factorization theorem, which we wish to seize in the following.

The purpose of this work is to use our new FF's for charged pions and kaons
\cite{bkk2} to make predictions for inclusive single-charged-particle
production in $\gamma\gamma$ collisions,
which can soon be confronted with first experimental data from LEP2.
With the exception of inclusive $\pi^0$ production \cite{gordon}, there exist
no NLO predictions for inclusive particle production at LEP2 in the literature.
We also consider $\gamma\gamma$ physics at a next-generation $e^+e^-$ linear
collider (NLC) with CM energy $\sqrt S=500$~GeV.
Furthermore, we compare our NLO analysis with $\gamma\gamma$ data at low CM
energies measured some time ago with the TASSO detector at DESY PETRA
\cite{tasso} and with the MARK~II detector at SLAC PEP \cite{markii}.
These data are at rather low $p_T$, where our NLO formalism is not expected to
yield reliable results, so that a meaningful test of our FF's may not be
possible.

As is well known, three mechanisms contribute to the production of quarks and
gluons in $\gamma\gamma$ collisions: $(i)$ In the production through direct
photons (DD), the two photons directly couple to the quark lines in the
hard-scattering amplitudes.
At least to LO, no spectator particles travel along the photon axes.
$(ii)$ If one of the photons splits into a flux of quarks (and gluons), one of
these quarks may directly interact with the second photon. The remaining
quarks (and gluons) build up a spectator jet in the direction of the split
photon (single-resolved (DR) $\gamma$ contribution). The $\gamma\gamma$
cross section of this mechanism depends on the PDF's of the photon.
$(iii)$ If both photons split into quarks and gluons
(double-resolved (RR) $\gamma$ contribution), two spectator jets appear. Since
the PDF's of the photon take large values---at small $x$ through the gluon part
and at large $x$ through the so-called point-like quark part---the DR and RR
contributions are numerically of the same order as the DD one.
We perform a consistent NLO analysis, {\it i.e.}, we include the DD, DR, and
RR hard-scattering cross sections, the photon PDF's, and the FF's at NLO.

The outline of this work is as follows.
In Section~2, we shortly describe the formalism which we use to calculate the
DD, DR, and RR contributions up to NLO.
Section~3 contains our numerical results for PETRA, PEP, LEP2, and NLC
energies.
Here, we also compare with the TASSO and MARK~II data. 
Our conclusions are summarized in Section~4.

\section{Formalism}

When we speak of $\gamma\gamma$ collisions with almost real photons, we have
in mind the $e^+e^-$-collision process where the positrons and electrons
act as sources of nearly massless, collinear photons, which collide with each
other to produce a debris of hadrons in the final state. The energy spectrum
for the produced photons is well described by the equivalent-photon
approximation (EPA). In this approximation, the longitudinal- and
transverse-momentum components decouple. The transverse momentum is integrated
out with certain constraints, specified by the experimental situation. The
resulting spectra for the photons emitted from the positrons and electrons are
thus functions of the longitudinal-momentum fractions, $x_1$ and $x_2$, 
respectively.
These distributions depend on the experimental setup.
In our analysis, we consider past and future experiments
at the colliders PETRA, PEP, LEP2, and NLC (TESLA design or laser spectrum).
The respective forms of the EPA will be specified later when we present our
numerical results.

Our notation is as follows. We consider the reaction
\begin{equation}
\label{process}
e^+(p_1)+e^-(p_2)\to e^+(p_1')+e^-(p_2')+h(p_h)+X,
\end{equation}
where $h$ is the observed hadron and X includes all unobserved hadrons.
The four-momentum assignments are indicated in the parentheses.
In the EPA, the inclusive cross section of process~(\ref{process})
is related to that of the corresponding $\gamma\gamma$ reaction,
\begin{equation}
\label{not1}
\gamma_1(p_1^{\gamma})+\gamma_2(p_2^{\gamma})\to h(p_h)+X,
\end{equation}
through
\begin{eqnarray}
E_h\frac{d^3\sigma(e^+e^-\to e^+e^-h+X)}{d^3p_h}
&\n=\n&\int\limits_{x_1^{min}}^1dx_1
\int\limits_{x_2^{min}}^1dx_2\,F_{\gamma_1/e^+}(x_1)F_{\gamma_2/e^-}(x_2)
\nonumber\\
\label{gg}&\n\times\n&
E_h\frac{d^3\sigma(\gamma_1\gamma_2\to h+X)}{d^3p_h},
\end{eqnarray}
where $x_i=E_i^\gamma/E_i$ $(i=1,2)$ and
$F_{\gamma_i/e^{\pm}}(x_i)$ stand for the
photon-spectrum functions to be specified later on.
The lower bounds of integration, $x_{1,2}^{min}$, are fixed by kinematics in
terms of the transverse momentum $p_T$ and CM rapidity $y$ of $h$.
In the applications to follow, the actual $e^+$ and $e^-$ acceptances
must be incorporated in the integration ranges of $x_1$ and $x_2$,
respectively.

In the QCD-improved parton model, the cross section of reaction~(\ref{not1})
is expressed as a convolution of the parton-parton scattering cross sections,
where one or both of the incoming partons may be photons, with the
scale-dependent PDF's and FF's,
\begin{eqnarray}
\label{xgg}
E_h\frac{d^3\sigma (\gamma_1\gamma_2\to h+X)}{d^3p_h}
&\n=\n&\int\frac{dx_h}{x_h^2}\sum_cD_{h/c}(x_h,M_h^2)\left[
p_c^0\frac{d^3\sigma (\gamma_1\gamma_2\to c+X)}{d^3p_c}
 \right. \nonumber\\
&\n+\n&\sum_a\int dx_aF_{a/\gamma}(x_a,M_{\gamma}^2)
    p_c^0\frac{d^3\sigma(a\gamma_2\to c+X)}{d^3p_c}
 \nonumber\\
&\n+\n&\sum_b\int dx_bF_{b/\gamma}(x_b,M_{\gamma}^2)
    p_c^0\frac{d^3\sigma(\gamma_1b\to c+X)}{d^3p_c}
 \nonumber\\
&\n+\n&\sum_{a,b}\int dx_aF_{a/\gamma}(x_a,M_{\gamma}^2)
    \int dx_bF_{b/\gamma}(x_b,M_{\gamma}^2)
 \nonumber\\
&\n\times\n&\left.p_c^0\frac{d^3\sigma(ab\to c+X)}{d^3p_c}
\right].
\end{eqnarray}
Here, the parton indices $a,b,c$ run over the gluon and $N_F$ flavours of
quarks and antiquarks, $k_a=x_ap_1^{\gamma}$, $k_b=x_bp_2^{\gamma}$, and
$k_c=p_h/x_h$ are the parton momenta,
$F_{a/\gamma}(x_a,M_\gamma^2)$ is the PDF of parton $a$ inside the photon,
and $D_{h/c}(x_h,M_h^2)$ is the FF of parton $c$ into hadron $h$.
The factorization scales, $M_\gamma$ and $M_h$, will be specified later on.
The first term on the right-hand side of Eq.~(\ref{xgg}) is the direct-photon
contribution (DD), the second and third ones are the once-resolved 
contributions (DR), and the fourth one is the twice-resolved contribution (RR).
In LO, the DD contribution is of $O(\alpha^2)$, where $\alpha$ is the
electromagnetic coupling constant, {\it i.e.}, it is of the same order as the
the respective hard-scattering cross sections.
The LO hard-scattering cross sections in the DR and RR components are of
$O(\alpha\alpha_S)$ and $O(\alpha_S^2)$, respectively, so that, at first 
sight, the LO DR and RR contributions might be considered to be of higher
orders in the strong coupling constant, $\alpha_S$.
However, they are formally and numerically of the same order as the LO
direct part, since the quark PDF's of the photon are enhanced by the factor
$\alpha/\alpha_S$ and the gluon PDF is enhanced at small $x$ \cite{drees1}.
In fact, the factor $\alpha/\alpha_S$ from the photon PDF's properly adjusts
the orders of the DR and RR contributions.

It is convenient to define PDF's for finding a particular parton $a$,
with momentum fraction $x$, inside the positron or electron, 
$F_{a/e^{\pm}}(x,M^2)$. In the EPA, this function is given by the
convolution of the respective photon PDF, $F_{a/\gamma}(x,M^2)$, with
the photon-spectrum function, $F_{\gamma/e^{\pm}}(x)$, introduced in
Eq.~(\ref{gg}), viz.
\begin{equation}
\label{epdf}
F_{a/e^{\pm}}(x,M^2)=\int\limits_x^1\frac{dy}{y}\,
  F_{a/\gamma}\left(\frac{x}{y},M^2\right)F_{\gamma/e^{\pm}}(y).
\end{equation}
If the photon directly participates in the hard scattering, the
photon PDF in Eq.~(\ref{epdf}) must be replaced by the delta function
$\delta(1-x/y)$.
Using definition~(\ref{epdf}), we can combine Eqs.~(\ref{gg}) 
and (\ref{xgg}) in one
formula. The inclusive cross section to NLO is then written as
\begin{eqnarray}
\label{dsigma}
E_h \frac{d^3\sigma(e^+e^-\to e^+e^-h+X)}{d^3p_h}
&\n=\n&\sum_{a,b,c}\int dx_1dx_2\, 
   \frac{dx_h}{x_h^2}
   F_{a/e^{\pm}}(x_1,M_{\gamma}^2)F_{b/e^{\pm}}(x_2,M_{\gamma}^2)
   D_{h/c}(x_h,M_h^2)\nonumber\\
   &\n\times\n&\frac{1}{\pi s}\left[\frac{1}{v}\,
   \frac{d\sigma_{k_ak_b\to k_c}^0}{dv}(s,v;\mu^2)\delta(1-w)
   \right. \nonumber\\
   &\n+\n&\left.\frac{\alpha_S(\mu^2)}{2\pi}
   K_{k_ak_b\to k_c}(s,v,w;\mu^2,M_{\gamma}^2,M_h^2)\right],
\end{eqnarray}
where it is understood that the QCD-corrected hard-scattering cross sections
in the DD, RR, and two DR channels are properly multiplied with the 
respective $e^+$ and $e^-$ PDF's and summed over.
As usual, $v=1+t/s$ and $w=-u/(s+t)$, where $s=(k_a+k_b)^2$,
$t=(k_a-k_c)^2$, and $u=(k_b-k_c)^2$ are the Mandelstam variables at the parton
level.
They are related to the external Mandelstam variables, $S=(p_1+p_2)^2$,
$T=(p_1-p_2)^2$, and $U=(p_2-p_h)^2$, by $s=x_1x_2S$, $t=x_1T/x_h$, and
$u=x_2U/x_h$.
The functions $K_{k_ak_b\to k_c}(s,v,w;\mu^2,M_{\gamma}^2,M_h^2)$ contain the
NLO corrections to the hard-scattering cross sections, and $\mu$ is the QCD
renormalization scale.

The $K_{k_ak_b\to k_c}$ functions for the DD channel have been derived by
Aurenche {\it et al.} \cite{aurenche1} and have recently been confirmed by
Gordon \cite{gordon}. The DR $K_{k_ak_b\to k_c}$ functions have been
calculated in Ref.~\cite{aurenche2} and reevaluated in Ref.~\cite{gordon},
where explicit expressions may be found.
The RR $K_{k_ak_b\to k_c}$ functions have been obtained in Ref.~\cite{aversa}.
They have recently been applied to predict
single-charged-particle production in low-$Q^2$ $ep$ scattering \cite{kk}.
Direct and resolved photoproduction in $ep$ scattering
correspond to the DR and RR components in $\gamma\gamma$ scattering, and we
may convert our previous analysis by replacing the proton PDF's with the
photon PDF's.
For the evaluation of the DD contribution, we employ a computer code created
by Gordon in connection with Ref.~\cite{gordon}.

For consistency, one needs both the photon PDF's and FF's in NLO.
As photon PDF's we use the NLO set by Gl\"uck, Reya, and Vogt (GRV) \cite{grv},
which we translate into the $\overline{\mbox{MS}}$ factorization scheme.
In their analysis, the
$M_{\gamma}^2$~evolution starts at a rather low value, $M_0^2=0.3$~GeV$^2$.
Therefore, this set is also applicable at rather small $p_T$. However, one
should keep in mind that the predictions may not be reliable in the
small-$p_T$ region because $\alpha_S$ is large and nonperturbative effects may
dominate. As FF's for charged particles, {\it i.e.}, the sum of charged pions
and kaons, we employ our recently constructed NLO set \cite{bkk2}, which has
been extracted in the $\overline{\rm MS}$ factorization scheme. 
This scheme has also been employed for the derivation of
the NLO kernels $K_{k_ak_b\to k_c}$.

In the calculation of the $K_{k_ak_b\to k_c}$ functions pertinent to the DD
and DR channels, one encounters collinear singularities associated with the
splitting of the incoming photons into $q{\bar q}$ pairs.
These initial-state singularities are absorbed into the bare PDF's appearing
in the resolved-photon contributions, {\it i.e.}, the DR and RR components,
which renders these functions $M_{\gamma}$ dependent. A similar $M_{\gamma}$
dependence, but with opposite sign, shows up in the $K_{k_ak_b\to k_c}$
functions of the DD and DR processes. In this way, the
DD, DR, and RR processes become interrelated. The DR
(RR) part is actually a NLO contribution to the DD (DR) part that cannot be
treated fully perturbatively anymore. This
nonperturbative part is then described by the photon PDF's.
Up to higher-order terms in the photon PDF's, the $M_{\gamma}$ dependence
cancels in the combination of the DD and DR contributions on one hand,
and the DR and RR contributions on the other hand.
It is clear that the classification into DD, DR, and RR contributions becomes
ambiguous if NLO corrections are included.
NLO terms that have been attributed to the DD (DR) contribution may be shifted
to the DR (RR) part.
The cancellation of the dependences on $M_{\gamma}$ and the choice of
factorization scheme associated with the incoming-photon leg in the
superposition of direct and resolved photoproduction in $ep$ scattering
has been demonstrated in Ref.~\cite{kk}.
We expect that this cancellation also works for the $\gamma\gamma$
process~(\ref{not1}).
The formal aspects of such cancellations in $\gamma\gamma$ reactions have been
investigated by Gordon \cite{gordon}.

\section{Numerical Results}

We are now in a position to present our numerical results
for the cross section of inclusive charged-particle production in
$\gamma\gamma$ collisions at NLO.
Our plan is as follows.
First, we consider existing data taken at low energies
by TASSO \cite{tasso} and MARK~II \cite{markii} and
confront them with the corresponding NLO calculations.
Then, we make predictions for LEP2 and NLC, superimposing the energy spectra
of the usual EPA bremsstrahlung off the incoming positrons and electrons and
of beamstrahlung according to the TESLA design to be discussed later.
As a further application, we consider $\gamma\gamma$ collisions with photon
spectra as they result from Compton backscattering of laser light off
the $e^+$ and $e^-$ beams of the NLC.
If this method works,
it will be possible to generate high-luminosity beams of real photons carrying
approximately 80\% of the $e^+$ ($e^-$) energy (see Ref.~\cite{drees2}
for the impact of various NLC collider options).

We work in the $\overline{\rm MS}$ renormalization and factorization
scheme with $N_F=5$ active quark flavours and
$\Lambda_{\overline{\rm MS}}^{(5)}=$~131~MeV, except for our TASSO and MARK~II
analyses, where we use $N_F=4$ and
$\Lambda_{\overline{\rm MS}}^{(4)}=$~200~MeV.
We identify the various scales and set $\mu=M_{\gamma}=M_h=\xi p_T$, where
$\xi$ is a dimensionless scale factor.
Unless stated otherwise, we put $\xi=1$ and employ the NLO
set of the GRV photon PDF's \cite{grv}
in the $\overline{\rm MS}$ scheme. 
We take the NLO FF's for charged pions and kaons from our recent work
\cite{bkk2}.
We sum over charged charged pions and kaons.
Charged-baryon production is likely to be negligible;
in $e^+e^-$ annihilation, some 10\% of the charged tracks are due to protons
and antiprotons \cite{bkk2}, and one expects their fraction to be even smaller
in $\gamma\gamma$ processes.
In our MARK~II analysis, we calculate the ratio of the $K_S^0$ and
charged-hadron cross sections.
For this purpose, we employ the $K_S^0$ FF's from our very recent work
\cite{bkk3}.

In the following three subsections, the results are ordered according to
the various energy regions relevant for PETRA/PEP, LEP2, and NLC.

\subsection{Results for PETRA/PEP Energies}

Before presenting our predictions at high energies (LEP2 and NLC), where we
expect to see experimental data at large $p_T$ in the near future---at least
from LEP2---, we wish to investigate how well the QCD-improved parton model
can explain existing data at lower energies. Unfortunately, there are 
only few data for the inclusive cross section with information on the
absolute normalization. Such data come from TASSO at PETRA \cite{tasso} and
MARK~II at PEP \cite{markii}. In the TASSO experiment, the effective
CM energy was $\sqrt{S}=33.1$~GeV, which is the average for runs with beam
energies between 13.7 and 18.3~GeV \cite{hilger}. The tagging requirements
for the positrons and electrons were such that one of the two leptons (positron
or electron) was tagged in the forward detector, which covered a narrow angular
region between $\theta_{min}=24$~mrad and $\theta_{max}=60$~mrad relative to
the beam direction, if it carried an energy of at least 4~GeV, whereas the
other lepton (electron or positron) remained untagged \cite{tasso}.
In our theoretical
evaluation, we take these tagging conditions into account by calculating the
energy spectrum of the photons from the tagged leptons by means of the formula
\begin{equation}
\label{ftasso}
F_{\gamma/e}(x)=\frac{\alpha}{2\pi}\left[\frac{1+(1-x)^2}{x}
   \ln\frac{Q_{max}^2}{Q_{min}^2}
+2m_e^2x\left(\frac{1}{Q_{max}^2}-\frac{1}{Q_{min}^2}\right)\right],
\end{equation}
where $m_e$ is the electron mass, $Q_{min}^2=(1-x)E^2\theta_{min}^2$, and
similarly for $Q_{max}^2$ ($E=\sqrt{S}/2$ is the beam energy).
The tag-energy condition leads to a photon-energy cutoff at $x_{max}=0.76$.
The no-tag photon spectrum is calculated from the well-known
formula by Brodsky {\it et al.} \cite{brodsky}, 
written in a convenient form in Ref.~\cite{kni}.
Furthermore, we include an overall factor of 2 to account
for the fact that both the positron or the electron can trigger the forward
detector.
The average $Q^2$ of the tagged photon was $\langle Q^2\rangle=0.35$~GeV$^2$
\cite{tasso}.
In our evaluation, we take $Q^2=0$ in the hard-scattering cross sections and
photon PDF's.
This is justified since $\langle Q^2\rangle\ll p_T^2$ for $p_T\simgt2$~GeV,
which we are primarily interested in.
A further experimental constraint in the TASSO
experiment was $|\cos{\theta}|<0.84$, where $\theta$ is the angle of
charged tracks with respect to the beam axis \cite{tasso}.
This restricts the $y$ range over which we integrate the doubly
differential cross section $d^2\sigma/dy\,dp_T^2$ to be $|y|<1.22$.
Taking these kinematical constraints into account, we
calculate $d\sigma/dp_T^2$ for charged particles. 

The result is shown in Fig.~\ref{fig1} as a function of $p_T$ for the three 
different scale choices $\xi=1/2$, 1, 2 and compared with the TASSO data.
As we can see, the experimental cross section is systematically larger than the
theoretical one.
In particular at larger $p_T$, where our approach is supposed to be valid,
the measurement greatly exceeds our prediction, even for $\xi=1/2$.
At small $p_T$, we do expect that the data overshoot our
theoretical result, since in this region soft interactions give rise to
additional particle production, which has not been subtracted from the data.
This contribution is supposed to be significant for $p_T\simlt1.5$~GeV and
is usually described as in hadronic reactions induced by the
vector-dominance-mechanism (VDM) component of the photon.
We have no explanation for the disagreement with the data in the upper
$p_T$ range. A serious background for the production of high-$p_T$ events
from one-photon annihilation events, where either of the incoming positron or
electron radiates a hard photon, was thoroughly investigated by TASSO
\cite{tasso} and was subtracted from the data shown in Fig.~\ref{fig1}.
It is interesting to study the relative importance of the DD, DR, and RR 
contributions.
At $p_T\simlt1$~GeV, the RR component is dominant, whereas 
for larger $p_T$ the bulk of the cross section comes from the DD channel.
At $p_T=(2$, 3, 4)~GeV, the DD component amounts to (51, 69, 75)\%,
respectively.
This shows that, at these $p_T$ values, the use of different photon PDF's
would not change the cross section in any appreciable way and that the
disagreement with the data is due to the DD component.

Single-tag measurements similar to those by TASSO were
performed with the MARK~II detector at PEP \cite{markii} operating with
$\sqrt{S}=29$~GeV.
The virtualities of the photons emitted from the tagged lepton were required to
lie between $Q_{min}^2=0.075$~GeV$^2$ and $Q_{max}^2=1$~GeV$^2$ and had the
average value $\langle Q^2\rangle=0.5$~GeV$^2$ \cite{markii}.
The $x$ range of these photons and the $y$ range of the produced hadrons were
only constrained by kinematics \cite{markii}.
The MARK~II data are confronted with our theoretical predictions for 
$\xi=1/2$, 1, and 2 in Fig.~\ref{fig2}.
The theoretical curves are multiplied by an overall factor 1.2, to account for
$p_T$ smearing and resolution effects not corrected for in the experimental
data \cite{slac}.
We observe that the agreement between theory and experiment is considerably
better than in the case of TASSO.
At rather low $p_T$, where theory is not expected to be reliable
because of the importance of the soft hadronic component, which was not 
subtracted in the data, there is almost perfect agreement, while the
theoretical curves undershoot the data at large $p_T$.
The dent at $p_T\approx3$~GeV in the curve for $\xi=1$ is due to the onset of
$c$-quark fragmentation \cite{bkk2}.
The disagreement between the MARK~II charged-particle cross section and the
theoretical predictions at larger $p_T$ has previously been noticed by Aurenche
{\it et al.} \cite{aurenche4} and by Gordon \cite{gordon}.
However, these authors used older FF's, partly in LO \cite{aurenche4}.
Our results, which are based on FF's that are rigorously constrained by recent
$e^+e^-$ data \cite{bkk2}, suggest that this disagreement does not originate
from the FF's.
It also appears that, at large $p_T$, our results exceed those of 
Refs.~\cite{gordon,aurenche4}.

The MARK~II Collaboration also measured the cross section of inclusive $K_S^0$
production under kinematic constraints identical to those used for their 
charged-hadron sample \cite{markii}.
Using our recently constructed set of $K_S^0$ FF's \cite{bkk3},
we calculate the ratio of $K_S^0$ to charged hadron production as a function of
$p_T$.
The results are compared with the MARK~II data in Fig.~\ref{fig3},
showing satisfactory agreement for $p_T\simgt1$~GeV.
The step in the theoretical curve at $p_T\approx3$~GeV comes about because the
$c$-quark threshold affects charged-hadron and $K_S^0$ production somewhat
differently.

\subsection{Results for LEP2}

Unfortunately, there are no LEP1 data available which could be directly
compared with our theoretical results.
ALEPH \cite{aleph} and DELPHI \cite{delphi} reported on inclusive measurements
of charged tracks, extending up to $p_T\approx4$~GeV.
However, these are not absolutely normalized, and the EPA constraints, which
are indispensable for absolute predictions, are not specified.

It is hoped that better data, which extend to even larger $p_T$, will become
available very soon from LEP2. For our predictions, we assume
$\sqrt{S}=175$~GeV and describe the quasireal-photon spectrum in the EPA
by the formula \cite{frix}
\begin{eqnarray}
\label{frix}
F_{\gamma/e}(x)&\n=\n&\frac{\alpha}{2\pi}\left\{\frac{1+(1-x)^2}{x}
  \ln{\frac{E^2\theta_{max}^2(1-x)^2+m_e^2x^2}{m_e^2x^2}}
  \right.\nonumber\\
  &\n+\n&\left.2(1-x)\left[\frac{m_e^2x}{E^2\theta_{max}^2(1-x)^2+m_e^2x^2}
  -\frac{1}{x}\right]\right\}.
\end{eqnarray}
We vary $x=E_{\gamma}/E_e$ over the full range allowed by kinematics and
put the antitagging angle to $\theta_{max}=30$~mrad.

In Fig.~\ref{fig4}, we show the DD, DR, and RR contributions to the cross
section $d\sigma/dp_T$ of inclusive charged-hadron production and their sum
as a function of $p_T$.
As in the case of MARK~II, $d\sigma/dp_T$ is obtained from
$d^2\sigma/dy\,dp_T$ by integrating over the full $y$ range.
For $p_T\simgt5$~GeV, the cross section is dominated
by the DD component, which is then bigger than the sum of the DR and RR 
contributions.
In general, the DR and RR components yield only a small fraction of
the total sum, except for very small $p_T$ ($p_T\simlt3$~GeV), where the RR
part dominates. 
The $y$ spectrum for $p_T=10$~GeV is plotted in Fig.~\ref{fig5}, where
the DD, DR, and RR contributions are again displayed along with their sum.
Due to the symmetric experimental setup, the curves are symmetric in $y$.
For $p_T=10$~GeV, the DD contribution dominates for all values of $y$.
It is of interest to know whether the resolved contributions can be used to
obtain information about the gluon PDF of the photon.
Since for large $p_T$ the DR and RR contributions are small, this should be
feasible only for small $p_T$.
To investigate this point, we repeat the computation of $d\sigma/dp_T$ putting
$F_{g/\gamma}(x)=0$ and plot the ratio of the outcome to the full result
in Fig.~\ref{fig6} (dashed line).
We observe that this ratio is below 80\% only if $p_T\simlt5$~GeV.
This means that this cross section is not a very powerful discriminator
to disentangle the gluon PDF of the photon. Almost the same pattern
is observed if the gluon FF is switched off (dot-dashed line).
The cross section $d\sigma/dp_T$ is sensitive to $D_{g/h^{\pm}}(x)$ only for
small $p_T$. This may be understood by observing that the gluon
FF falls off much more rapidly with increasing $x$ than the quark FF's.
In Fig.~\ref{fig6}, we also show the fraction of $d\sigma/dp_T$
originating from the fragmentation into $K^{\pm}$. It is rather large, 
ranging between 34\% and 40\% for $p_T$ between 5 and 30~GeV.

It is generally believed that the scale dependence of a parton-model 
calculation to a given order indicates the size of unknown higher-order
corrections and thus may be used to estimate the theoretical uncertainty of
the prediction.
This leads us to study in Fig.~\ref{fig7} the $\xi$ dependence of
$d\sigma/dp_T$ and its DD, DR, and RR parts for $p_T=10$~GeV.
This scale dependence is determined mostly by the DD component, which
dominates the cross section at $p_T=10$~GeV.
In the range $1/2<\xi<2$, the total sum varies by $\pm8\%$ relative to its
value at $\xi=1$.
The discontinuity at $\xi=0.95$ stems from the $b$-quark threshold.

\subsection{Results for NLC}

Next, we consider the predictions for the NLC with $\sqrt{S}=500$~GeV and
TESLA design. At the NLC,
photons are produced not only by bremsstrahlung, but also via
synchrotron radiation emitted by one of the colliding bunches in the
field of the opposing bunch \cite{barklow}. This phenomenon has been
termed beamstrahlung. The details of the beamstrahlung spectrum depend
crucially on the design and operation mode of the NLC. In our
study, we select the TESLA design, where the unwanted effects of 
beamstrahlung are reduced to a tolerable level. We coherently superimpose
the EPA and beamstrahlung spectra. The EPA spectrum is computed from 
Eq.~(\ref{frix}) with $\theta_{max}=175$~mrad and the
beamstrahlung spectrum from the expression given in Ref.~\cite{barklow}, with
parameters $\Upsilon_{eff}=0.039$ and $\sigma_2=0.5$~mm \cite{schulte}.
This combined spectrum has recently been used also in a
study of heavy-quark production in $\gamma\gamma$ collisions \cite{cacciari}. 
The $p_T$ distribution, integrated over the full $y$ range, is shown in
Fig.~\ref{fig8} together with its DD, DR, and RR components.
Apart from an overall enhancement due to the increased EPA logarithm and the
additional beamstrahlung-induced contribution,
the $p_T$ spectrum looks very similar to the LEP2 case in Fig.~\ref{fig4}.
The corresponding $y$ spectra for $p_T=10$~GeV are shown in Fig.~\ref{fig9}.
Due to the admixture of beamstrahlung, their shapes differ somewhat from those
of the pure EPA case in Fig.~\ref{fig5}.

The highest possible photon energies with large enough luminosity may be
achieved by converting the NLC into a $\gamma\gamma$ collider via
backscattering of high-energetic laser light off the $e^+$ and $e^-$ beams
\cite{ginzburg}.
The resulting photon spectrum is given by
\cite{ginzburg}
\begin{equation}
\label{gins}
F_{\gamma}(x)=\frac{1}{G(\kappa)}\left[1-x+\frac{1}{1-x}-\frac{4x}
{\kappa(1-x)}+\frac{4x^2}{\kappa^2(1-x)^2}\right],
\end{equation}
where
\begin{equation}
G(\kappa)=\left(1-\frac{4}{\kappa}-\frac{8}{\kappa^2}\right)
  \ln{(1+\kappa)}+\frac{1}{2}+\frac{8}{\kappa}-\frac{1}{2(1+\kappa)^2}.
\end{equation}
This spectrum extends up to $x_{max}=\kappa/(1+\kappa)$. 
In our calculation we choose $\kappa=4.83$,
so that $x_{max}=0.83$.
If the experimental setup was arranged so that $\kappa>4.83$, $e^+e^-$ pairs
would be produced in the collisions of the primary laser photons and the
high-energetic backscattered photons.
In Fig.~\ref{fig10}, we present the DD, DR, and RR contributions to the
$p_T$ spectrum $d\sigma/dp_T$ together with their sum. 
Compared to the previous cases, the relative magnitudes of the DD,
DR, and RR components have completely changed.
At small $p_T$, at about 5~GeV say, the RR component is by far the largest,
whereas the DD component is negligible.
At large $p_T$, around 20~GeV, the relation is quite different;
the DD, DR, and RR contributions all have the same order of magnitude.
A similar situation was encountered in the analogous study of heavy-quark
production \cite{cacciari}.
Consequently, in the lower $p_T$ range, the cross section $d\sigma/dp_T$ of
inclusive particle production with a Compton collider
will serve as a powerful tool to extract information on the photon PDF's.
The corresponding $y$ spectra for $p_T=10$~GeV are plotted in
Fig.~\ref{fig11}. As far as the relative importance of the various components
is concerned, we recognize a pattern similar to Fig.~\ref{fig10}, which
refers to the integrated rapidity spectrum, at $p_T=10$~GeV.
On the other hand, the $y$ spectra are quite different in shape and relative
magnitude from those found in the TESLA case.
This may be understood by observing that the laser-photon spectrum is peaked
at the upper edge.
The RR component is now dominant and rather flat;
the DR component is somewhat smaller and exhibits a characteristic two-hump
structure with a significant depletion in the central region;
the DD component is greatly suppressed, but exhibits a shape similar to the
TESLA case.
As a consequence, there is a pronounced plateau in the $y$ spectrum of the
total sum.
Inspired by the observations made in Fig.~\ref{fig10}, we quantitatively study
in Fig.~\ref{fig12} the influence of the gluon PDF of the photon.
The dashed line represents the ratio of $d\sigma/dp_T$ with the gluon PDF
switched off to the full result as a function of $p_T$.
We see that, at $p_T=(5$, 15, 25)~GeV, (75, 42, 23)\% of the cross section
stems from the gluon PDF.
The NLO photon PDF set by Gordon and Storrow (GS) \cite{gs} differs from the
GRV set \cite{grv} mainly in the gluon PDF.
The ratio of $d\sigma/dp_T$ evaluated with the GS set to the GRV
result is visualized in Fig.~\ref{fig12} by the solid line, which is a smooth
interpolation.
Obviously, the GS and GRV predictions appreciably differ at small $p_T$,
by up to a factor of 2.
Thus, the NLC operated as a $\gamma\gamma$ collider would provide an excellent
laboratory to pin down the poorly known gluon PDF of the photon.
In Fig.~\ref{fig12}, we also assess the sensitivity of such an experiment to
the gluon FF by showing the fraction of $d\sigma/dp_T$ due to the quark 
FF's (dot-dashed line).
At $p_T=(5$, 15, 25)~GeV, (45, 21, 13)\% of the cross section is related to
gluon fragmentation.
Consequently, the $\gamma\gamma$ NLC would also allow one to probe the gluon
FF, which is only moderately constrained by data of inclusive hadron
production via $e^+e^-$ annihilation.

\section{Summary and Conclusions}

We studied the inclusive production of charged hadrons in collisions of
almost real photons at NLO in the framework of the QCD-improved parton model.
This approach is conceptually very different from the one based on MC event
generators, which is frequently followed by experimentalists to interpret
their data.
In these MC programs, the various QCD processes are simulated using LO matrix 
elements in connection with certain model assumptions concerning the formation
of hadronic final states.
Although these MC packages often lead to satisfactory descriptions of the
data, from the theoretical point of view, their drawback is that this happens
at the expense of introducing a number of ad-hoc fine-tuning parameters,
which do not originate in the QCD Lagrangian.
Furthermore, in the MC approach, it seems impossible to implement 
the factorization of final-state collinear singularities, which impedes a 
consistent extension to NLO.
On the contrary, in the QCD-improved parton model, such singularities are 
absorbed into the bare (infinite) FF's so as to render them renormalized
(finite) in a way quite similar to the procedure for the PDF's at the incoming
legs.
In our analysis, we used FF's extracted with NLO precision from PEP and recent
LEP1 data on inclusive single-hadron production in $e^+e^-$ annihilation
\cite{bkk2,bkk3}.
In fact, owing to the factorization theorem \cite{facth}, the FF's are process
independent as long as one considers unbiased single-hadron event samples.
By the same token, our formalism is less flexible than the MC approach, since
we are unable to describe more complicated final states, with more than one
detected hadron and acceptance cuts on observables different from our
integration variables.

There are various ways to realize $\gamma\gamma$ processes in experiments.
The conventional method is to exploit the QED bremsstrahlung which is radiated
during $e^+e^-$ reactions from the colliding beams.
High-quality data of this type were collected by TASSO \cite{tasso} at PETRA
and MARK~II \cite{markii} at PEP.
Unfortunately, the bulk of these data are accumulated at rather small $p_T$,
where the NLO formalism is not expected to yield reliable results, and the
few data points at larger $p_T$ values carry considerable error bars due to
limited statistics.
Detailed comparison revealed that our consistent NLO calculation with
up-to-date FF's was able to reduce somewhat the mismatch between theory and
experiment that had previously been observed by other authors
\cite{gordon,aurenche4}.
However, especially in the case of TASSO, the degree of discrepancy in the
upper $p_T$ range still gives some reason for concern.

While at LEP1 the $\gamma\gamma$ processes were overwhelmed beneath the 
tremendous background due to $e^+e^-$ annihilation on the $Z$ resonance,
the situation will be much more favourable at LEP2.
We presented NLO predictions for the $p_T$ and $y$ spectra to be measured in
the oncoming LEP2 phase.
We verified that our NLO predictions are rather stable under scale variations;
at $p_T=10$~GeV, the cross section only fluctuates by $\pm8\%$ if $\xi$ is 
varied between 1/2 and 2.
Unfortunately, except at rather low $p_T$, the study of $\gamma\gamma$
processes at LEP2 will not much deepen our understanding of the nature of the
gluon inside the photon and of its r\^ole within the fragmentation process.

Photon-photon physics at the NLC will greatly benefit from beamstrahlung as an
additional source of quasireal photons and, of course, from the increased EPA
logarithm.
In the case of the 500-GeV NLC with TESLA architecture, the cross sections at
$p_T=(5$, 15, 25)~GeV exceed the corresponding LEP2 values by factors of
(19, 26, 37), respectively, while the shapes of the $p_T$ and $y$ spectra
closely resemble their LEP2 counterparts.

The advent of a Compton collider, generated by backscattering of laser light
off the $e^+$ and $e^-$ beams of the NLC, would represent an important step
forward in the history of $\gamma\gamma$ experimentation, and might
considerably improve our knowledge of both the gluon FF and the gluon PDF of
the photon.
By converting the NLC, the cross sections of inclusive charged-hadron 
production at $p_T=(5$, 15, 25)~GeV would be increased by factors of
(35, 52, 88), respectively.
At the same time, the relative importance of the DD channel would be 
dramatically reduced, so that the sensitivity to the photon PDF's would be
correspondingly amplified.

\bigskip
\centerline{\bf ACKNOWLEDGMENTS}
\smallskip\noindent

We thank E. Hilger and N. Wermes for very useful communications concerning
Ref.~\cite{tasso}, G. Gidal for drawing our attention to Ref.~\cite{slac},
and L.E. Gordon for providing us with the FORTRAN implementation of his NLO
results for the DD channel \cite{gordon}.
One of us (G.K.) is grateful to the Theory Group of the
Werner-Heisenberg-Institut for the hospitality extended to him during a visit
when this paper was prepared.

\newpage

\vskip-6cm

\begin{figure}

\centerline{\bf FIGURE CAPTIONS}

\caption{\protect\label{fig1} Cross section $d\sigma/dp_T^2$ of inclusive 
charged-hadron production in single-tagged $\gamma\gamma$ collisions.
The TASSO data \protect\cite{tasso} ($\protect\sqrt{S}=33.1$~GeV on average)
are compared with our corresponding NLO calculations for scale choices
$\xi=1/2$, 1, and 2.
\hskip9cm}

\vskip-.2cm

\caption{\protect\label{fig2} Cross section $d\sigma/dp_T$ of inclusive 
charged-hadron production in single-tagged $\gamma\gamma$ collisions.
The MARK~II data \protect\cite{markii} ($\protect\sqrt{S}=29$~GeV)
are compared with our corresponding NLO calculations for scale choices
$\xi=1/2$, 1, and 2.
\hskip9cm}

\vskip-.2cm

\caption{\protect\label{fig3} The ratio of inclusive $K_S^0$ production to
inclusive charged-hadron production as measured by MARK~II 
\protect\cite{markii} is compared with our NLO calculation for scale choice
$\xi=1$.
The rise near $p_T=3$~GeV is related to the $c$-quark threshold.
\hskip5cm}

\vskip-.2cm

\caption{\protect\label{fig4} NLO cross section $d\sigma/dp_T$ of inclusive 
charged-hadron production in double-tagged $\gamma\gamma$ collisions at LEP2
($\protect\sqrt{S}=175$~GeV).
The DD, DR, and RR components are also shown.
\hskip5cm}

\vskip-.2cm

\caption{\protect\label{fig5} Same as in Fig.~\protect\ref{fig4} for the $y$ 
dependence of $d^2\sigma/dy\,dp_T$ at $p_T=10$~GeV.
\hskip5cm}

\vskip-.2cm

\caption{\protect\label{fig6} Influence of the gluon PDF of the photon,
the gluon FF, and the charged-kaon final states on the $\xi=1$ result of
Fig.~\protect\ref{fig4}.
Shown are the fractions that remain if the gluon is switched off in the photon
PDF's (dashed line) or in the FF's (dot-dashed line) as well as the fraction
due to charged-kaon production (solid line).
\hskip5cm}

\vskip-.2cm

\caption{\protect\label{fig7} $\xi$ dependence of the results shown in
Fig.~\protect\ref{fig4} at $p_T=10$~GeV, in the interval $0.3<\xi<4$.
\hskip5cm}

\vskip-.2cm

\caption{\protect\label{fig8} Same as in Fig.~\protect\ref{fig4} for the NLC
with TESLA design ($\protect\sqrt{S}=500$~GeV).
\hskip5cm}

\vskip-.2cm

\caption{\protect\label{fig9} Same as in Fig.~\protect\ref{fig5} for the NLC
with TESLA design ($\protect\sqrt{S}=500$~GeV).
\hskip5cm}

\vskip-.2cm

\caption{\protect\label{fig10} Same as in Fig.~\protect\ref{fig4} for the NLC
in the Compton-backscattering mode ($\protect\sqrt{S}=500$~GeV).
\hskip5cm}

\vskip-.2cm

\caption{\protect\label{fig11} Same as in Fig.~\protect\ref{fig5} for the NLC
in the Compton-backscattering mode ($\protect\sqrt{S}=500$~GeV).
\hskip5cm}

\vskip-.2cm

\caption{\protect\label{fig12} Influence of the gluon PDF of the photon and
the gluon FF on the total result of Fig.~\protect\ref{fig10}.
Shown are the fractions that remain if the gluon is switched off in the photon
PDF's (dashed line) or in the FF's (dot-dashed line) as well as the ratio of
the calculation with the GS photon PDF's to that with the GRV set (solid line).
The solid line is a smooth interpolation.
\hskip5cm}

\end{figure}

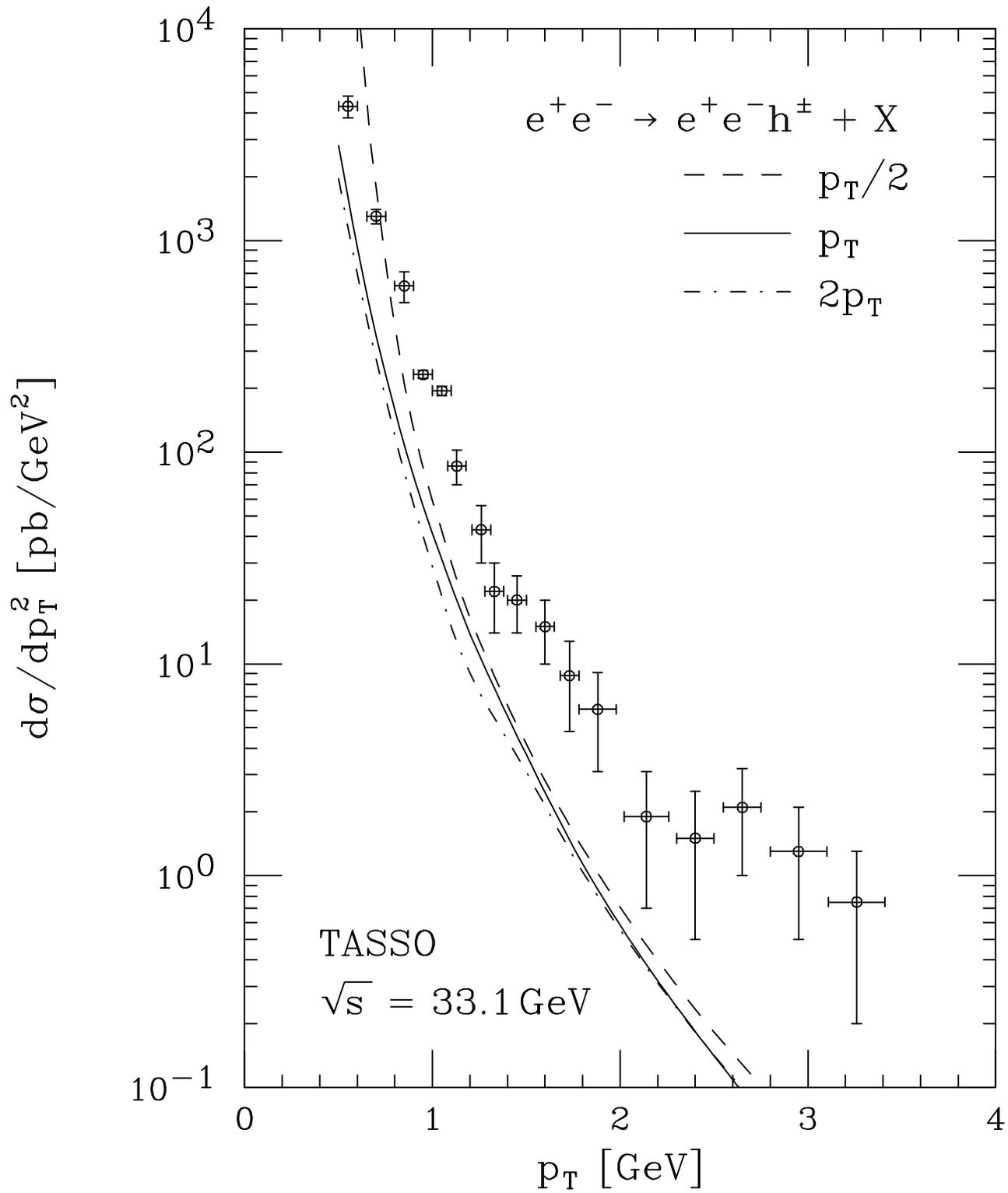
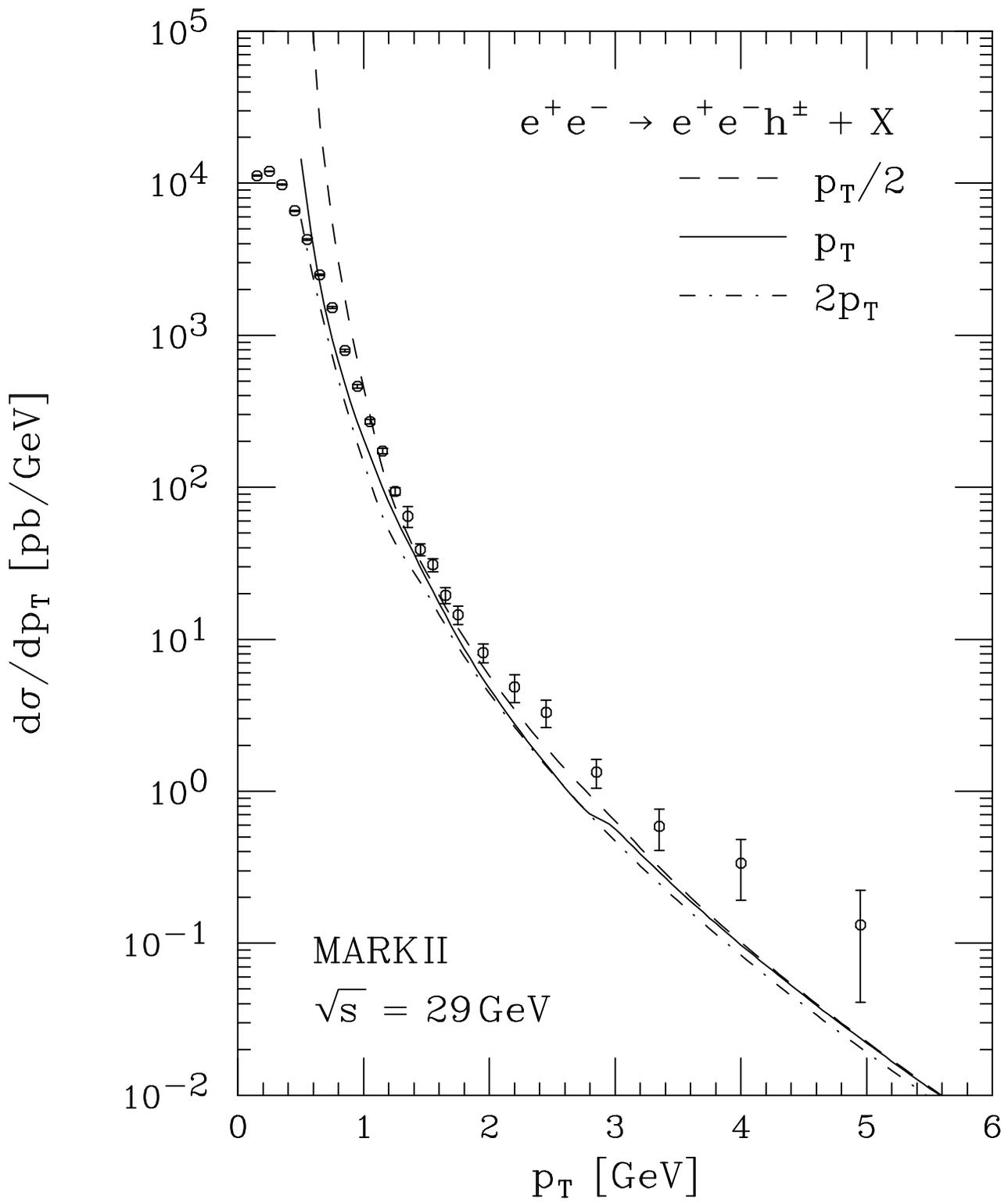
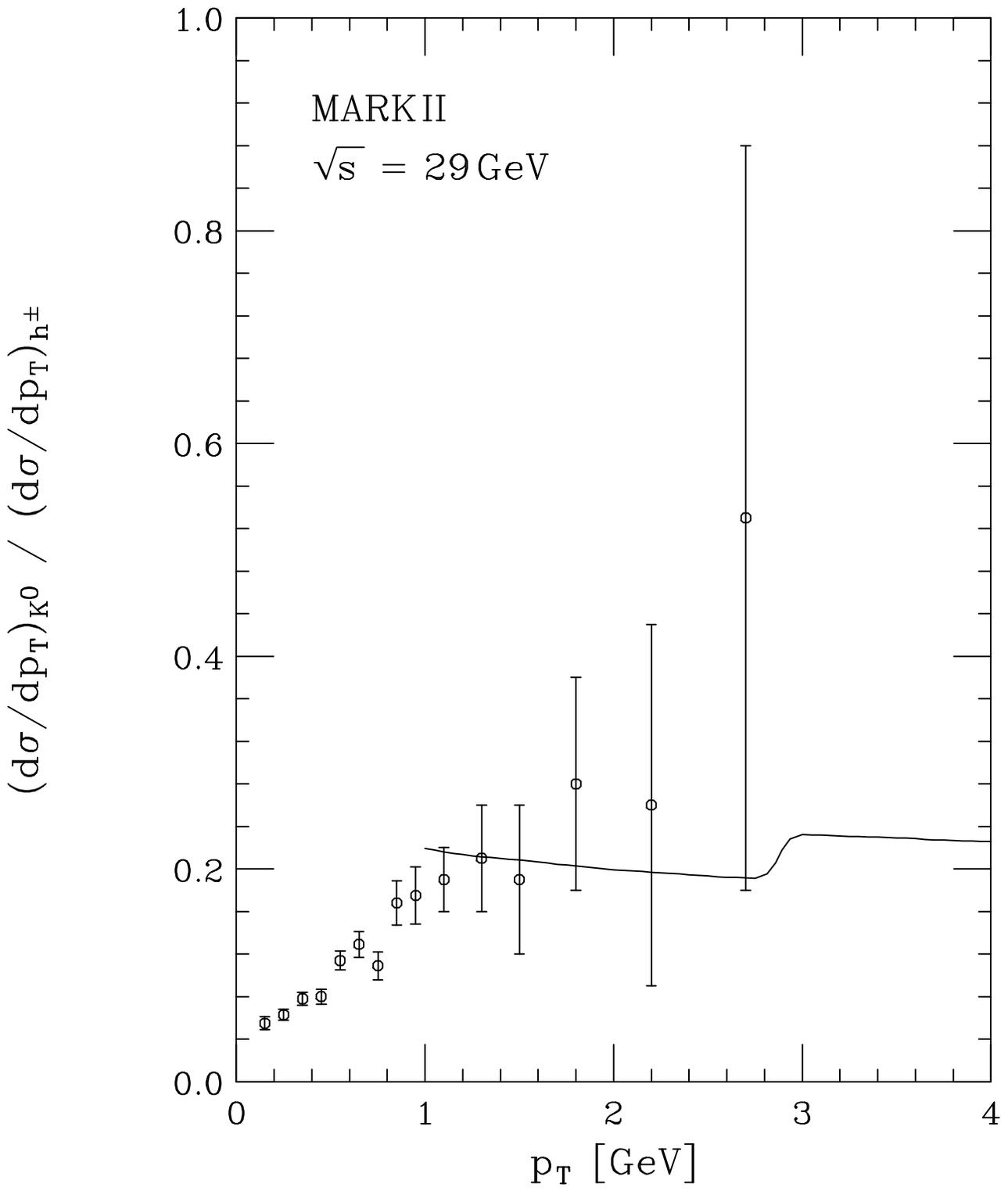
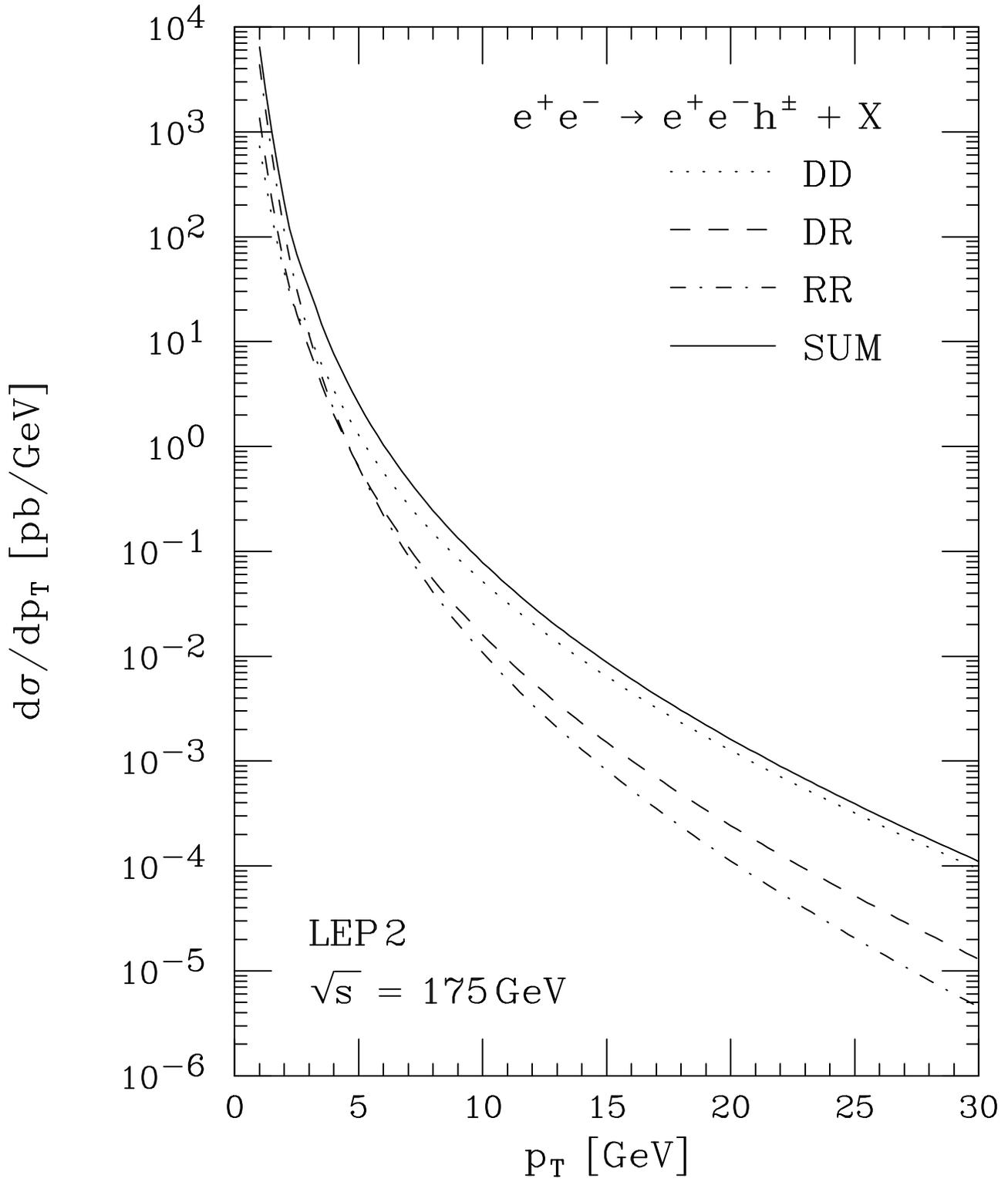
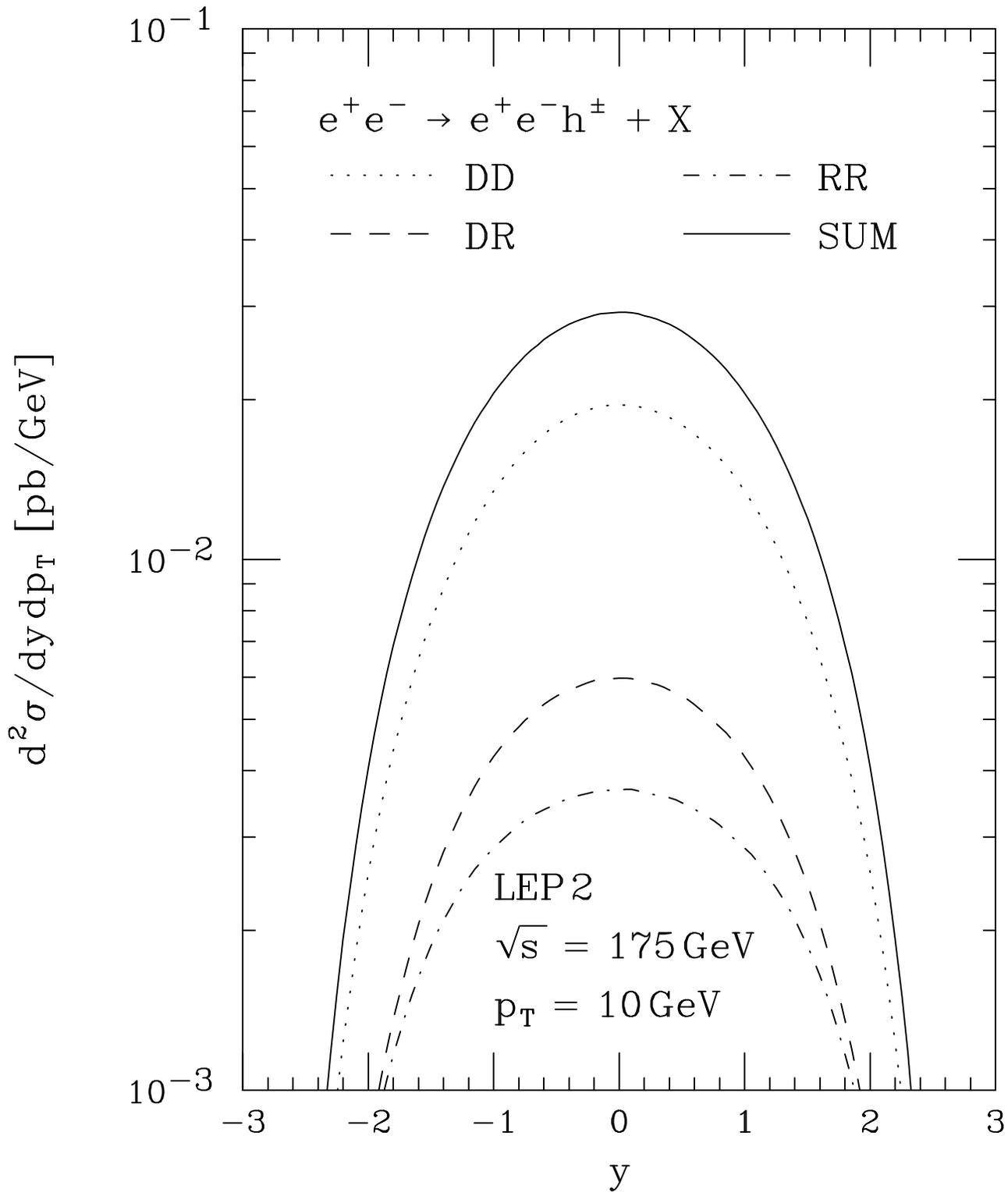
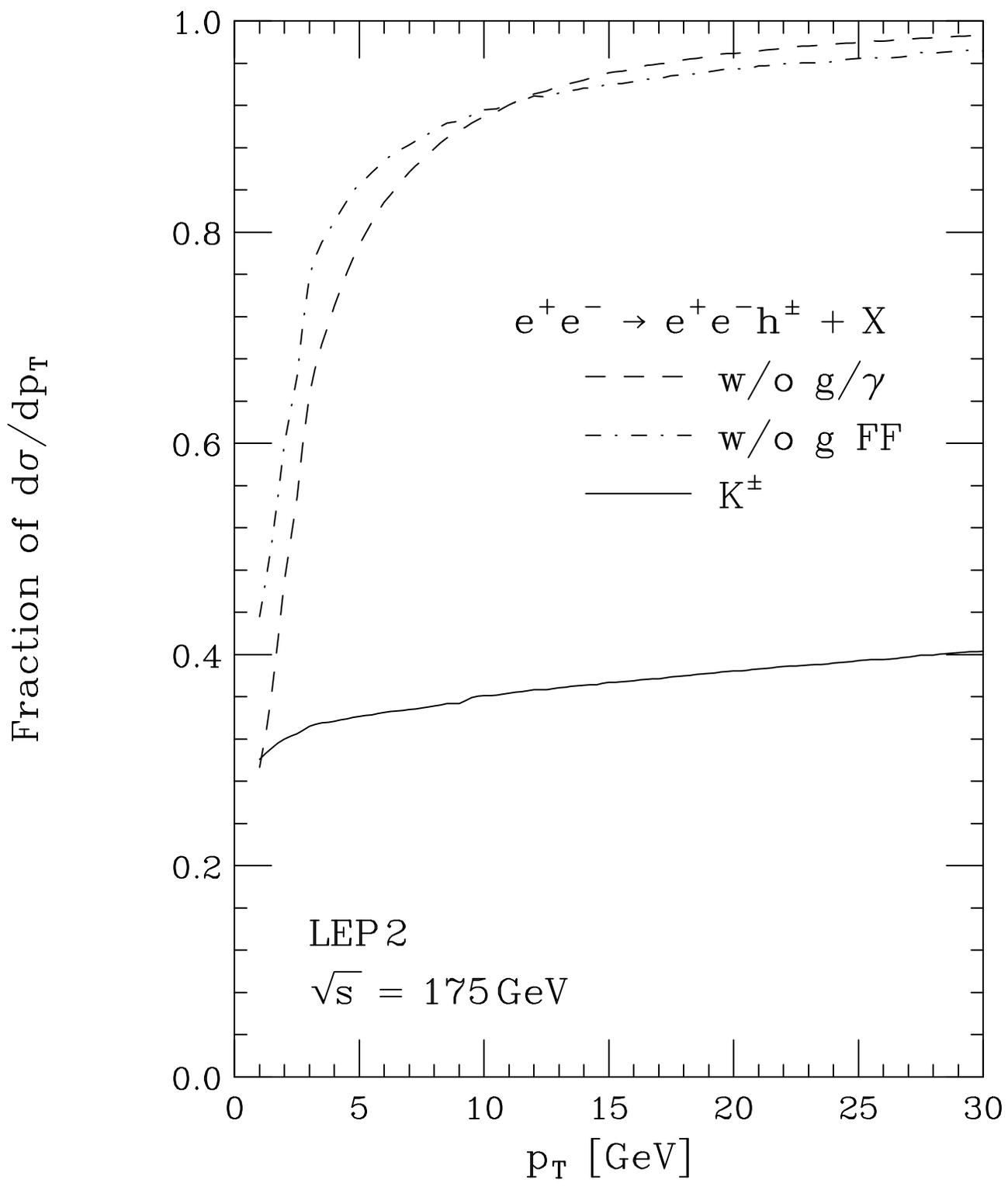
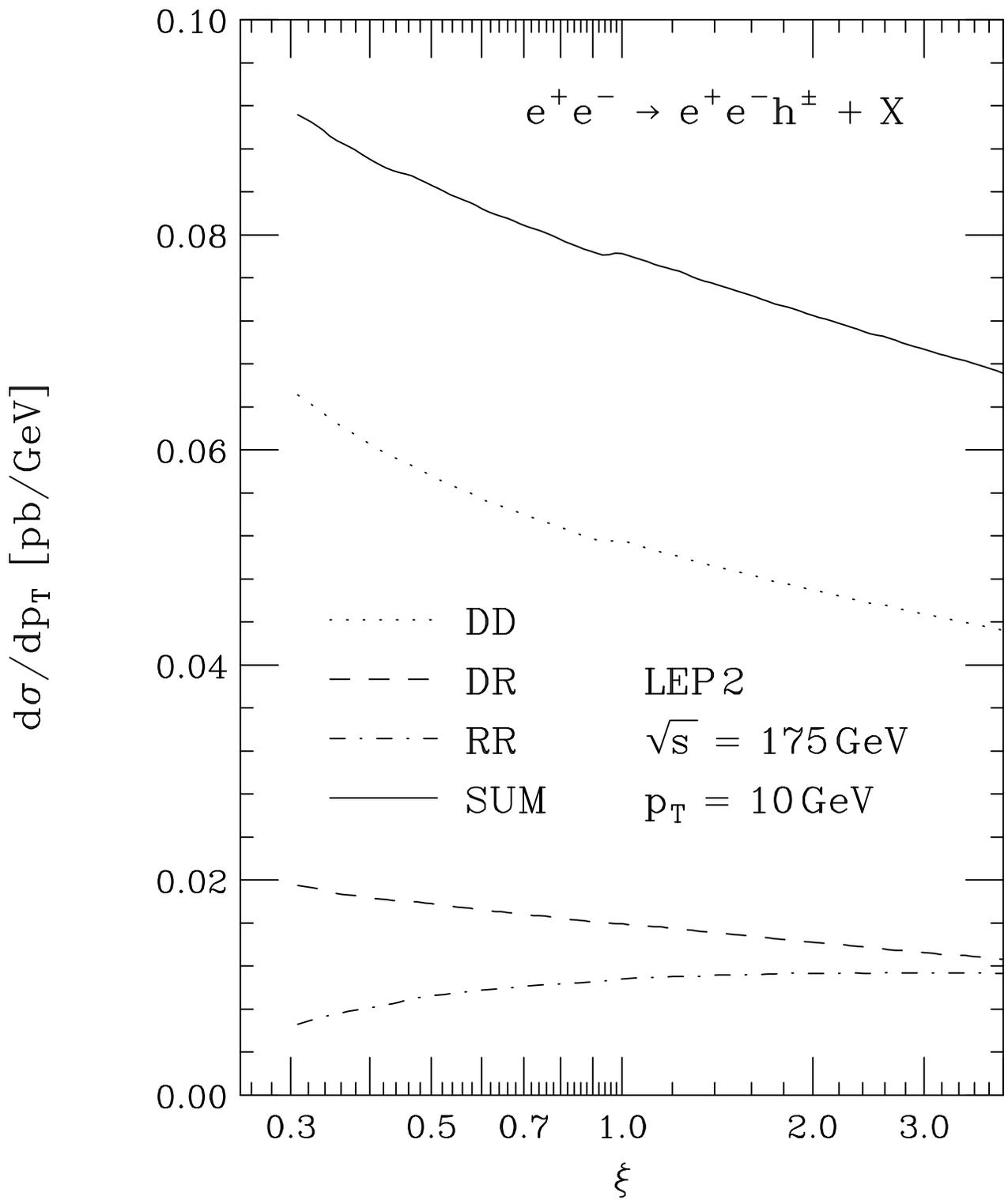
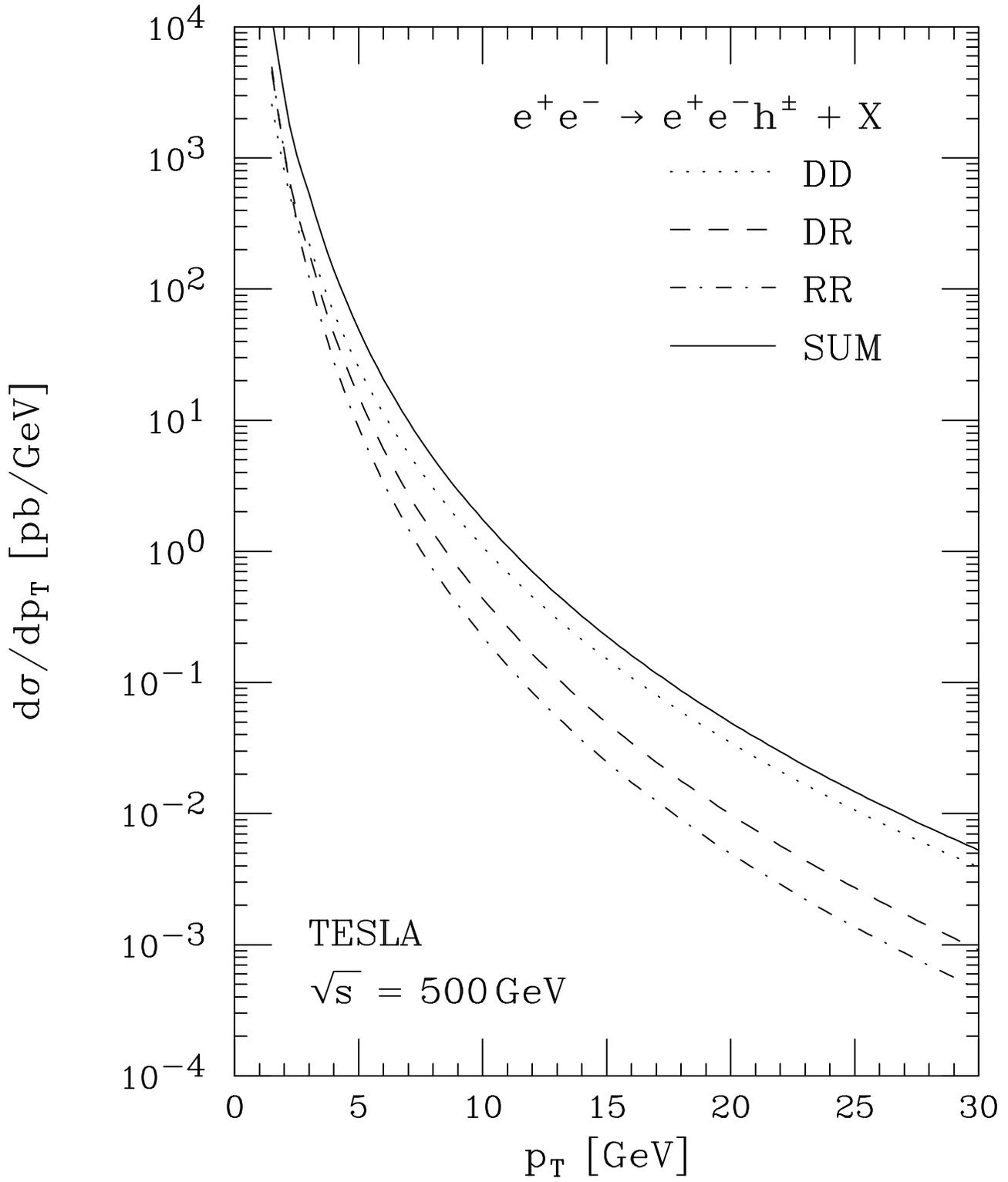
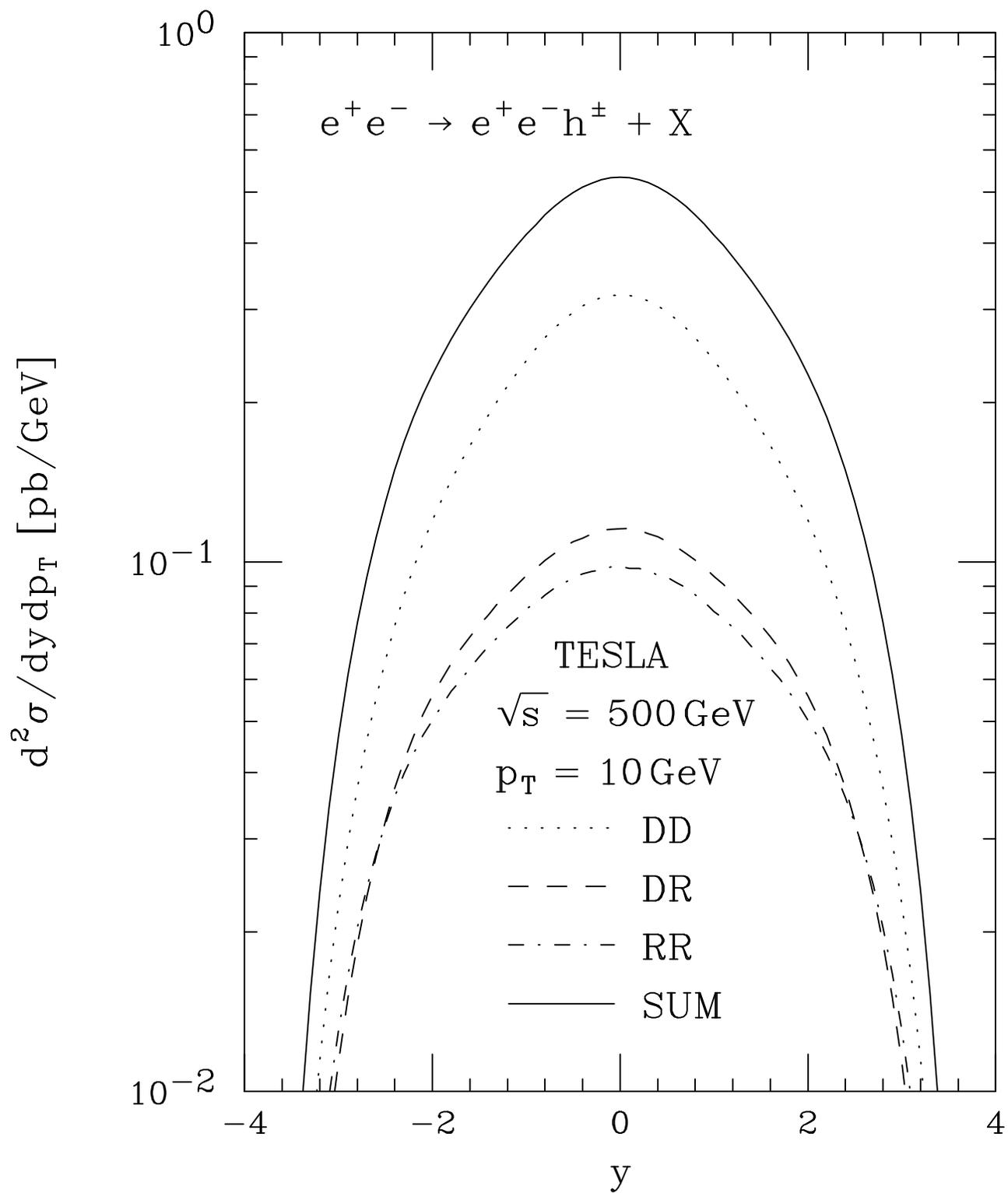
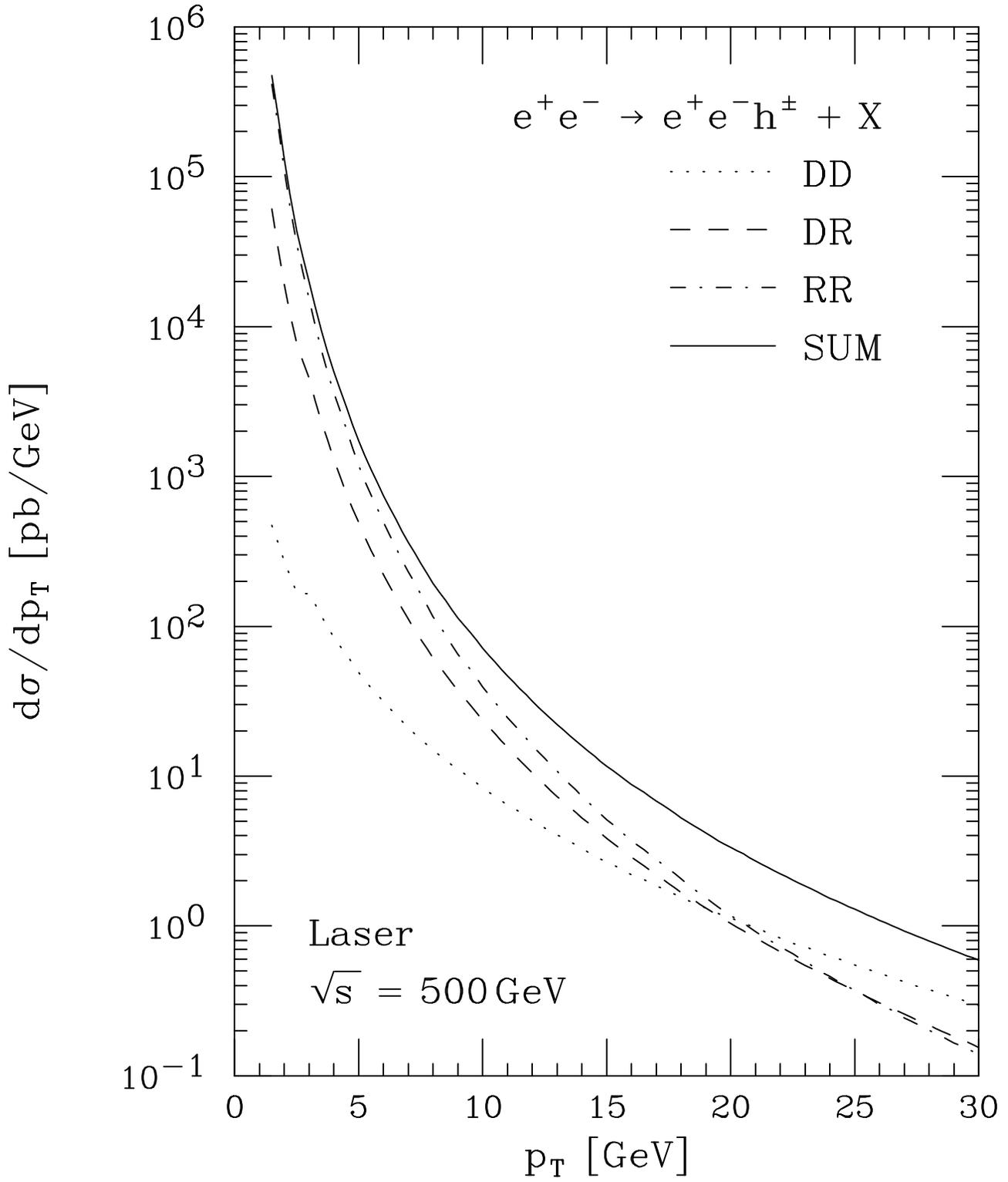
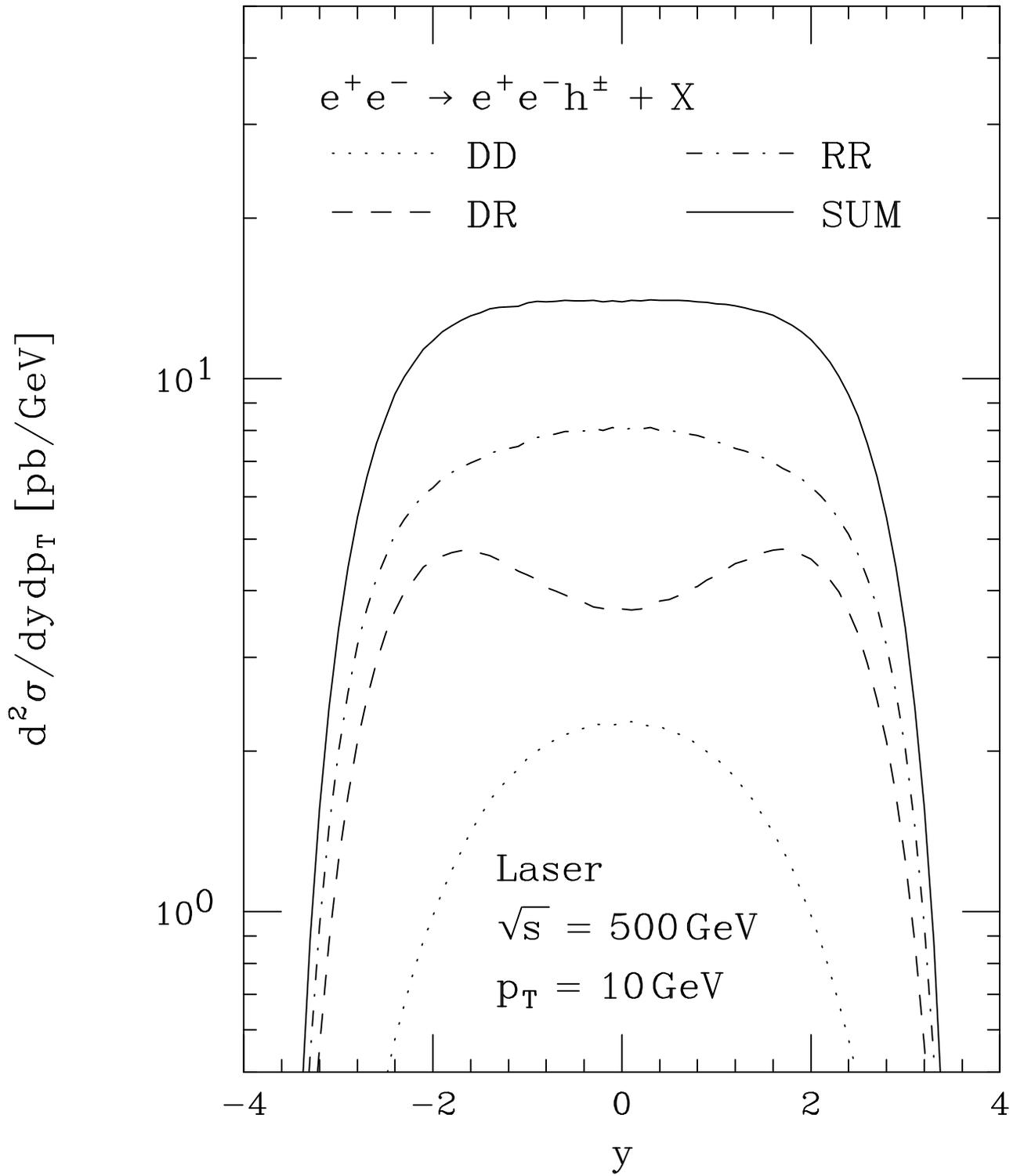
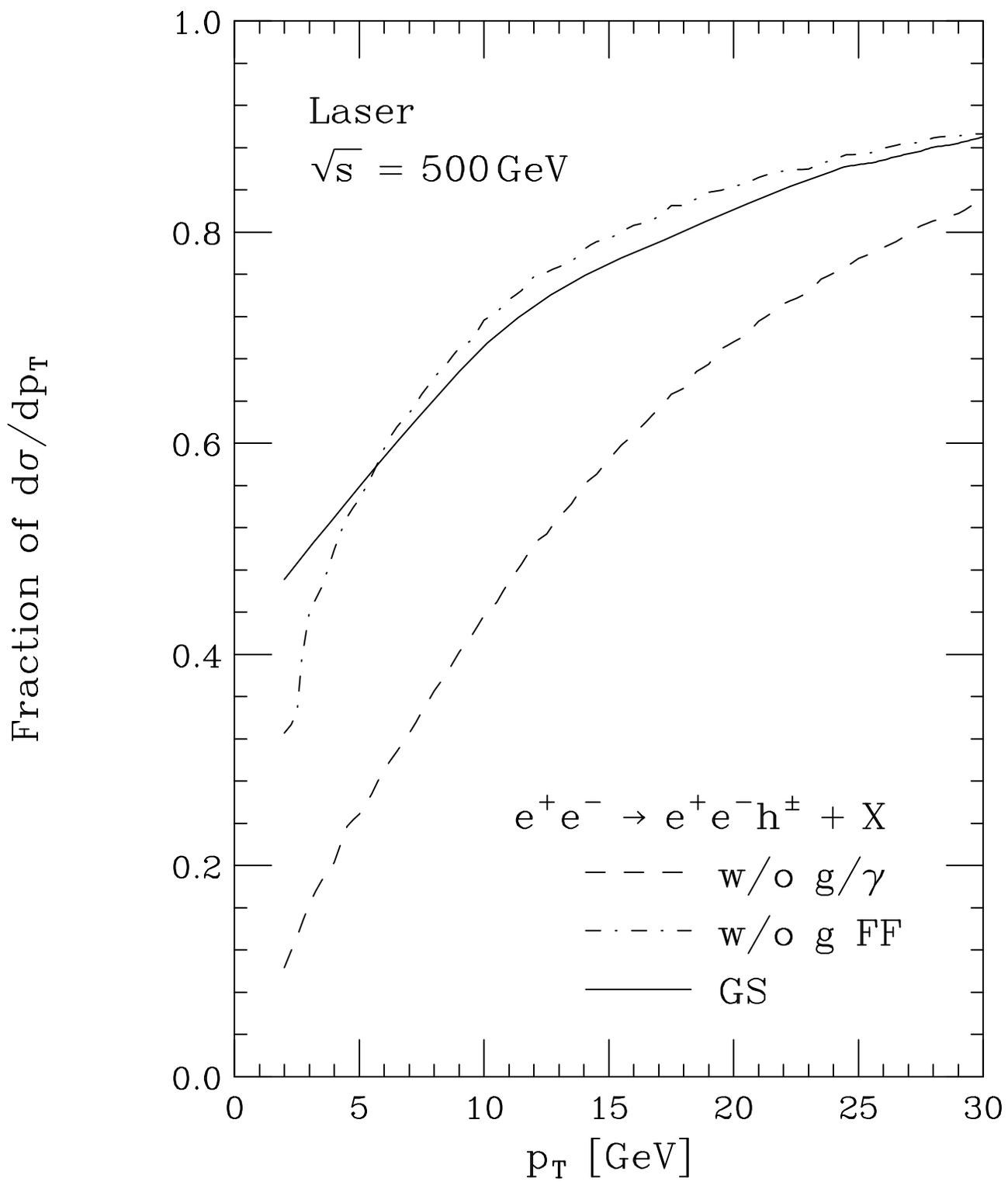
\end{document}